\documentclass{llncs}

\usepackage{times}
\usepackage{epsfig}
\usepackage{graphicx}
\usepackage{amsmath}
\usepackage{amssymb}
\usepackage{algorithm}
\usepackage{algpseudocode}
\usepackage{float}
\usepackage{array}
\usepackage{makecell}
\begin{document}

\title{Fixed-MAML for Few Shot Classification in \\Multilingual Speech Emotion Recognition}

\author{Anugunj Naman, Chetan Sinha, Liliana Mancini}
\institute{Indian Institute of Information Technology, Guwahati \\ Cardiff University, UK \\ \{anugunj.naman, chentan.sinha\}@iiitg.ac.in \\ liliana.mancini@cardiff.ac.uk}

\maketitle

\begin{abstract}
In this paper, we analyze the feasibility of applying few-shot learning to speech emotion recognition task (SER). The current speech emotion recognition models work exceptionally well but fail when then input is multilingual. Moreover, when training such models, the models' performance is suitable only when the training corpus is vast. This availability of a big training corpus is a significant problem when choosing a language that is not much popular or obscure. We attempt to solve this challenge of multilingualism and lack of available data by turning this problem into a few-shot learning problem. We suggest relaxing the assumption that all N classes in an N-way K-shot problem be new and define an N+F way problem where N and F are the number of emotion classes and predefined fixed emotion classes, respectively. We propose this modification to the Model-Agnostic Meta Learning (MAML) algorithm to solve the problem and call this new model F-MAML. This modification performs better than the original MAML and outperforms on EmoFilm dataset. 

\end{abstract}

\section{Introduction}\label{sec:Introduction}

Emotion recognition plays a significant role in many intelligent interfaces \cite{piccard:17}. Even with the recent advances in machine learning, this is still a challenging task. The main reason behind this is that most publicly available annotated datasets in this domain are small in scale, which makes DL models prone to over-fitting. Another essential feature of emotion recognition is the inherent multi-modality in expressing emotions \cite{jengli:19}. Emotional information can be captured by studying many modalities, including facial expressions, body postures, and EEG \cite{nicu:1}. Of these, arguably, speech is the most accessible. In addition to accessibility, speech signals contain many other emotional cues \cite{Kim:2}. We, therefore, use speech signals as a base to predict the emotion.

Generally, in speech emotion recognition (SER) task, conventional supervised learning solves the problem efficiently given sufficient training data. Several studies on SER for different single corpora have been conducted using the language-dependent optimal acoustic sets over several decades. Such systems can be analyzed in monolingual scenarios; changing the source corpus requires re-selecting the optimal acoustic features and re-training the system. Human-emotion perception, however, has proved to be cross-lingual, even without the understanding of the language used \cite{Li:16}. An SER system is expected to recognize emotions as such.

However, for an automatic SER system to recognize emotion, there are two significant problems. First, the training corpus available for many different languages is very limited. Second, it is not clear which standard features are efficient in detecting emotions across different cultures. Commonalities and differences in human-emotion perception across languages in the valence-activation (V-A) space have recently been studied \cite{Li:16}. It was revealed that direction and distance from neutral to other emotions are similar across languages, and languages' neutral positions are language-dependent. In this paper, motivated by the above challenges, we want to simulate a scenario where one can provide few labeled speech samples in any language and train a model on that language for few iterations to get a robust SER system. This proposed scenario removes the requirement for a large amount of data and identifies the standard features efficient in detecting emotions about that culture and fine-tune to it accordingly. 

Supervised learning has been extremely successful in computer vision, speech, or machine translation tasks, thanks to improvements in optimization technology, larger datasets, and streamlined designs of deep convolutional or recurrent architectures. Despite these successes, this learning setup does not cover many aspects where learning is possible and desirable. One such instance is learning from very few examples in the so-called few-shot learning tasks \cite{garcia:2018}. Rather than depending on regularization to compensate for the lack of training data, researchers have explored ways to leverage the distribution of similar tasks, inspired by human learning \cite{Lake:1332}. A lot of useful solutions have been developed, and the most popular solution right now uses meta-learning.

Meanwhile, most of the studies on few-shot learning are conducted on image tasks. We here attempt to apply those meta-learning solutions to SER systems. We formulate the problem mentioned above as a few-shot learning problem and analyze the performance state-of-art model-level few-shot learning algorithms.

Meta-learning, also known as 'learning to learn,' aims to make quick adaptation to new tasks with only a few examples. Recently many different meta-learning solutions have been proposed to solve the few-shot learning problems. All these solutions differ in the form of learning a shared metric \cite{Vin:3}\cite{Snell:4}\cite{Sung:5}\cite{Ko:6}, a generic inference network \cite{San:7}\cite{Mishra:8}, a shared optimization algorithm \cite{Munkhdalai:9}\cite{Ravi:10}, or a shared initialization for the model parameters \cite{finn:11}\cite{zli:12}\cite{nichol:13}. In this paper, we use the Model-Agnostic Meta-Learning (MAML) approach \cite{finn:11} because of the following  reasons:

\begin{enumerate}
  \item It is a model-agnostic general framework that can be easily used on a new task.
  \item It achieves state-of-the-art performance in existing few-shot learning tasks.
\end{enumerate}

Few-shot learning is often defined as an N-way, K-shot problem where N is the number of class labels in the target task, and K is the number of examples of each class. In most previous studies, it is assumed that all the N classes or labels are new. However, in real-life applications, these classes or labels are not necessary to be all new. Thus, we further define an N+F-way, K-shot problem where N and F are the numbers of new classes or labels and fixed classes, respectively. In this new devised task, the model has to classify among new classes and fixed classes. We propose this modification to the original MAML algorithm to solve this problem and call this new model F-MAML.

We conduct our experiment on EmoFilm dataset \cite{Parada:14} to simulate a scenario in SER. We compare our approach with two baseline approaches: the conventional supervised learning approach\cite{Lim:244} and the MAML\cite{ramit:43} approach. Experimental results show that F-MAML lead to obvious improvement over the supervised learning approach and even performing better than MAML. Our contributions in this paper are summarized here:

\begin{enumerate}
  \item We analyze the feasibility of few-shot learning for training SER models.
  \item We propose an efficient way compared to MAML (F-MAML) to train future SER models for any language with few training examples.
\end{enumerate}

The rest of the paper is presented in the following manner: In section 2, we discuss our work's background. In section 3, we discuss our proposed method. In section 4, experiments performed in detail, and results are mentioned. In section 5, we finally give a conclusion.

\section{Background}\label{sec:Background}

In this section, we first briefly introduce MAML, the base, and our solution's motivation. 

\begin{figure}
\begin{center}
\includegraphics[width=5cm]{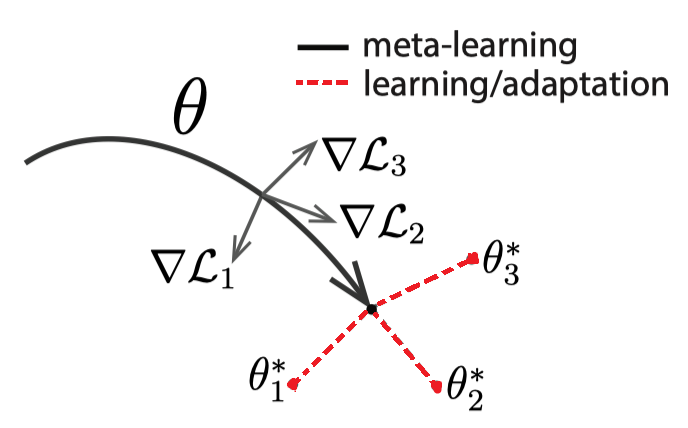}    
\caption{The MAML algorithm learns a good parameter initializer $\theta^*$ by training across various meta-tasks such that it
can adapt quickly to new tasks. Adapted from \cite{finn:11}.} 
\label{fig:mamll}
\end{center}
\end{figure}

Model-Agnostic Meta-Learning (MAML) is one of the most popular meta-learning algorithms that aim to solve the few-shot learning problem. The main goal of MAML is to train a model initializer that can adapt to any new task using very few labeled examples and training iterations \cite{finn:11}. The model is trained across several tasks to reach this goal, and it treats the entire task as a training example. The model is required to face different tasks to get used to adapting to new tasks. In this section, we describe the MAML training framework. As is shown in Figure-\ref{fig:mamll}, the optimization procedure consists of two stages. A meta-learning stage on the training data and a fine-tuning stage on the
testing tasks.

\subsection{The meta-learning stage}
Given that the target evaluation task is an N-way, K-shot task, the model is trained across a set of task ${T}$ where each task ${T_{i}}$ is also an N-way, K-shot task. In each iteration, a learning task, i.e., the meta-task ${T_{i}}$ is sampled according to a distribution over tasks ${p(T)}$. Each ${T_{i}}$ consist of a support set ${S_{i}}$ and a query set ${Q_{i}}$.

Consider a model represented by a parametrized function ${f_{\theta}}$ with parameters ${\theta}$.  ${\theta_{i}^{'}}$ is computed from ${\theta}$ through the adaptation to task ${T_{i}}$. A loss function ${\mathcal{L}_{S_{i}}(f_{\theta})}$, which is cross-entropy loss over support set examples, is defined to guide the computation of ${\theta_{i}^{'}}$:

\begin{equation} 
\label{eq:one}
\mathcal{L}_{S_{i}}(f_{\theta}) = - \sum_{(x_{j}, y_{j}) \in S_{i}}^{} y_{j}\log f_{\theta}(x_{j}).
\end{equation}

A one-step gradient update is as below:

\begin{equation} 
\label{eq:two}
{\theta_{i}^{'}} = \theta - \alpha\nabla_{\theta}\mathcal{L}_{S_{i}}(f_{\theta}).
\end{equation}

Here, ${\alpha}$ is the learning rate, which can be a fixed hyperparameter or learned like the Meta-SGD \cite{zli:12}. The gradient here is updated for multiple steps. 

After this, the model parameters are optimized on the performance of ${f_{\theta_{i}^{'}}}$ evaluated by the query set ${Q_{i}}$ with respect to ${\theta}$. ${\mathcal{L}_{Q_{i}}(f_{\theta_{i}^{'}})}$ is another cross entropy loss over query set examples:

\begin{equation} 
\label{eq:three}
\mathcal{L}_{Q_{i}}(f_{\theta_{i}^{'}}) = -\sum_{(x_{u}^{'}, y_{u}^{'}) \in Q_{i}}^{} y_{u}^{'}\log f_{\theta^{'}}(x_{u}^{'}).
\end{equation}

Broadly speaking, MAML aims to optimize the model parameters such that few gradient steps on a new task will ultimately lead to a maximally effective behavior on that new task. At the end of each training iteration, the parameters ${\theta}$ are updated as below:

\begin{equation} 
\label{eq:four}
{\theta} \leftarrow \theta - \beta\nabla_{\theta}\mathcal{L}_{Q_{i}}(f_{\theta_{i}^{'}}).
\end{equation}

Here, $\beta$ is the learning rate of the meta learner. To increase the stability of training , instead of only one task a batch of tasks is sampled in each iteration. The optimization is performed by averaging the loss across the tasks. Thus, equation (\ref{eq:four}) can be generalized to:

\begin{equation} 
\label{eq:five}
{\theta} \leftarrow \theta - \beta\nabla_{\theta} \sum_{i}^{} \mathcal{L}_{Q_{i}}(f_{\theta_{i}^{'}}).
\end{equation}

\subsection{The fine-tuning stage}
A fine-tuning is performed before the evaluation. In an N-way, K-shot task, K examples from each of the N class labels are available at this stage in the target task's support set. The model trained above in the meta-learning stage will now be fine-tuned according to equation (\ref{eq:two}) for a few iterations. The updated model will then be evaluated on the remaining unlabeled examples (the target task's query set).

\section{Proposed Method}\label{sec:Method}
In the original MAML, it is assumed that all class labels in the target task are new class labels. However, these class labels do not necessarily need to be all new.  In real-life applications, some of the class labels are known so that more examples of these class labels can be used in the meta-learning stage. This paper will call them fixed classes as we later fix their output positions in the neural network classifier. We call this task, which has to classify among new classes and fixed classes, an N+F-way, K-shot problem where N, F, K are the number of emotion class labels, fixed class labels, and  examples of each class for fine-tuning respectively. This problem of simultaneously classifying unseen and seen class labels has not been investigated in the original MAML. In our solution, we try to tackle the problem by proposing modifications to the MAML training framework. We believe that the N+F way, K-shot problem is more realistic and our modification to MAML applies to various tasks. We now describe our methodology for a few-shot SER task.

\begin{figure}
\begin{center}
\includegraphics[width=10cm]{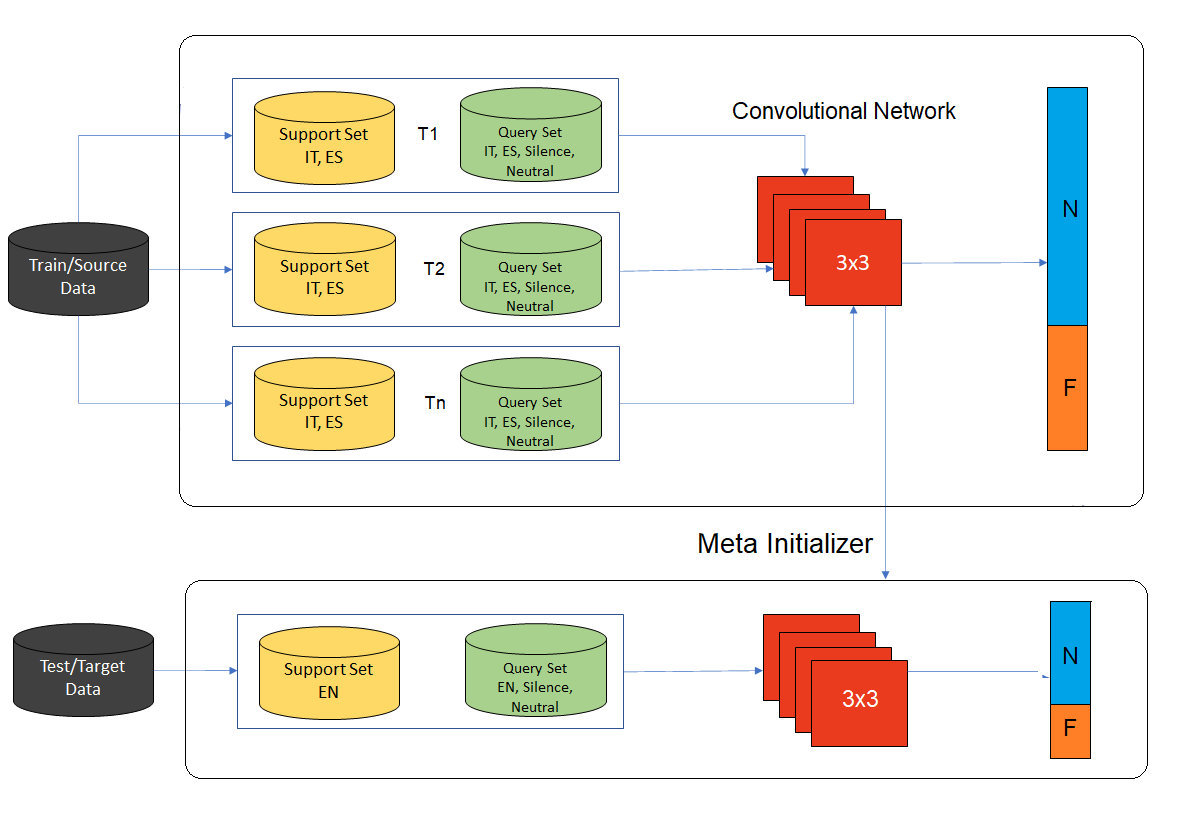}    
\caption{{Framework of our F-MAML approach for few-shot SER.}} 
\label{fig:framework}
\end{center}
\end{figure}

\subsection{Methodology: F-MAML}

Although the N+F-way, K-shot problem can be regarded as a specific form of the normal N-way, K-shot problem, solving it with the original MAML framework will lead to a performance degradation. Using the prior information of the F fixed classes, we modify the MAML framework in the following aspects:

\begin{enumerate}
  \item We fix the output positions, i.e., the output at the end of classification for a random sample for the fixed classes in the neural network classifier.
  \item These fixed classes occur in every meta-task $T_{i}$ in the meta-learning stage.
  \item The adaptation of fixed classes is not needed in the fine-tuning stage as they have already been learned in the meta-learning stage.
\end{enumerate}

The above three modification to the original MAML makes the proposed framework more effectively to real applications.

\subsection{Speech Emotion Recognition}

We formulate a scenario for SER as N+F-way, K-shot classification task. N is the number of emotions that one wishes to recognize, and one should provide K speech audio samples for each such emotion. Fixed labels here are silence and neutral. 

Figure \ref{fig:framework} illustrates the framework of F-MAML approach. The target data contains audio samples from one language, not in source data, while source data contain audio examples from all other languages. The fixed classes are the same in target and source data. In the meta-learning stage, several N+2-way, K-shot meta-task are sampled from source data for each language. Each meta-tasks is similar to the target task. We expect to learn a model initializer that can adapt to the target task using the provided speech samples and emotion labels. We exclude the fixed class labels from the support set in both meta-learning and the fine-tuning stages. As we can assume the availability of more training examples of fixed classes, we can keep them in the meta-tasks' query set in the meta-learning stage. Moreover, it can be seen that the positions of silence and the neutral classes are fixed to the last of network output (the orange area). Thus, we force our model to "recall" the fixed classes without the need for adaptation. 

\begin{algorithm}
\caption{MAML approach for few-shot ASR}\label{alg:xmaml}
\begin{algorithmic}[1]
\Require{$p(T)$ : distribution over tasks}
\Require{$X$ : training dataset}
\Require{$S_{il}$ : silence class set, $N_{eu}$: neutral class set}
\Require{$S_{i} \in X$ : support set, $Q_{i} \in [X \cup S_{il} \cup N_{eu}] \setminus S_{i}$ : query set}
\Require{$\alpha, \beta$: learning rates}
\State Randomly initialize base model parameters $\theta$
\While{not done}
\State {Sample a batch of meta-tasks $T_{i} \sim p(T)$}
\For{all $T_{i}$}{}
\State Sample a support set $S_{i} \in X$
\State Compute the gradient ${\mathcal{L}_{S_{i}}(f_{\theta})}$ using ${S_{i}}$ and $X$ as show in equation (\ref{eq:one}).
\State Update base model parameters with gradient descent: ${\theta_{i}^{'}} = \theta - \alpha\nabla_{\theta} \mathcal{L}_{S_{i}}(f_{\theta})$. (Step 6-7 can be repeated several times.)
\State Sample a query set $Q_{i}$ from the union ${[X, S_{il}, N_{eu}]} \setminus S_{i}$. (Selected emotion label from X in $Q_{i}$ and $S_{i}$ within $T_{i}$ are the same).
\State Compute the loss $\mathcal{L}_{Q_{i}}(f_{\theta_{i}^{'}})$ using $Q_{i}$ and the updated
model $f_{\theta^{'}}$.
\EndFor
\State Update the parameters for $\theta$ using each $Q_{i}$ and $\mathcal{L}_{Q_{i}}(f_{\theta^{'}})$:  ${\theta} \leftarrow \theta - \beta\nabla_{\theta} \sum_{i}^{} \mathcal{L}_{Q_{i}}(f_{\theta_{i}^{'}})$.
\EndWhile
\end{algorithmic}
\end{algorithm}

Algorithm \ref{alg:xmaml} summarizes the details of our approach. The algorithm described here is based on the work of \cite{finn:11} but is different in terms of how sampling is done for the support set and the query set during the meta-learning stage, which is introduced in section 3.1.

\section{Experimentation}\label{sec:Experiment}

\subsection{Dataset}
We conduct our experiments on EmoFilm dataset \cite{Parada:14}. It consists of 1115 clips with a mean length of 3.5 seconds, resulting in 341 English audio clips, with an average of 34.3 utterances per emotion; 410 Italian audio clips with an average of 41.3 utterances per emotion; and 356 Spanish clips, with an average of 35.9 utterances per emotion (std 9). The higher number of Italian clips might be due to Italian being a more 'emotionally expressive' language; this could also relate to the pre-test made by Italian listeners, who may be better at perceiving emotions in their language \cite{Parada:14}. The dataset is categorized into five emotion labels: happiness, sadness, anger, fear, and disgust. We formulate three 5-way, K-shot tasks using the same setup as the audio recognition tutorial in official PyTorch documentation. Table-\ref{tab:table1} gives information about total samples for each emotion in each language. We perform three experiments here. 

\begin{enumerate}
    \item The first experiment is SER in the English language, where we use the English language as a testing set while Spanish and Italian are used in training. 
    \item The second experiment is SER in the Italian language, where we use the Italian language as a testing set while English and Spanish are used in training.
    \item The third experiment is SER in the Spanish language, where we use the Spanish language as a testing set while English and Italian are used in training. 
\end{enumerate}

The testing language is unseen to the meta-learning stage, and only K labeled examples of each label are available in the fine-tuning stage. The initialized model is fine-tuned on the labeled examples and evaluated on the unlabeled examples. The samples for silence class and neutral class were self-generated with a mean length of 3.5 seconds. 

\begin{table}
\caption{Dataset Details}
\begin{center}
\vspace{-5mm}
\begin{tabular}{  c | c | c }
\thead{Language} & \thead{Total Samples} & \thead{Samples per Emotion} \\
\hline
English &  341 & \makecell{72 - Fear, 50 - Disgust, \\ 69 - Happiness, 76 - Anger, \\ 74 - Sadness} \\
\hline
Italian &  410 & \makecell{83 - Fear, 68 - Disgust, \\ 93 - Happiness, 73 - Anger, \\ 93 - Sadness} \\
\hline
Spanish &  356 & \makecell{63 - Fear, 50 - Disgust, \\ 76 - Happiness, 82 - Anger, \\ 85 - Sadness} \\
\hline
\end{tabular}
\end{center}
\label{tab:table1}
\end{table}

\subsection{Model Setting}

The 3-4 second clips are sampled at 16kHz. We use the Mel-frequency Cepstral Coefficient (MFCC) features. For each clip, we extract 40-dimensional MFCCs with a frame length of 30ms and a frame step of 10ms. Convolution Neural Networks is adopted as the base model, which contains  4 convolutional blocks. Each block comprises a 3 × 3 convolutions and 64 filters, followed by ReLU and batch normalization \cite{ioffe:16}. The flattened layer after the convolutional blocks contains 576 neurons and is fully connected to the output layer with a linear function. We avoided using ResNet architecture because it overfitted very quickly. The model is trained with a mini-batch size of 16 for 5, 10, 20-shot classification. We set the learning rate $\alpha$ to 0.1 and $\beta$ to 0.001. The learning rates were found using a grid search.

\subsection{Baselines}

We compare our proposed approach with two baseline approaches the state of art conventional supervised learning approach\cite{Lim:244}  which trains the model on the support set of the target task only, and the state of art meta learning approach MetaSER \cite{ramit:43}, which treats the 5+2-way problem as a 7-way problem. In the evaluation, we sample K examples from each class for fine-tuning the model and 25 examples per label for evaluation. We do 100 times random tests and evaluate different approaches on accuracy.

\subsection{Results}

We compare our approach with three baselines. Table \ref{tb:5shot}, \ref{tb:10shot} and \ref{tb:20shot} list the performance of 5, 10 and 20-shot task on SER in English, Spanish, Italian languages respectively. 

Not surprisingly, MetaSER, i.e., MAML based approaches perform much better than conventional supervised learning in a few-shot learning situation. This improvement is because it provides a good initialization of a model's parameters to achieve optimal fast learning on a new task with few gradient steps while avoiding overfitting that may happen when using a small dataset.

\begin{figure}
\begin{center}
\includegraphics[width=12cm]{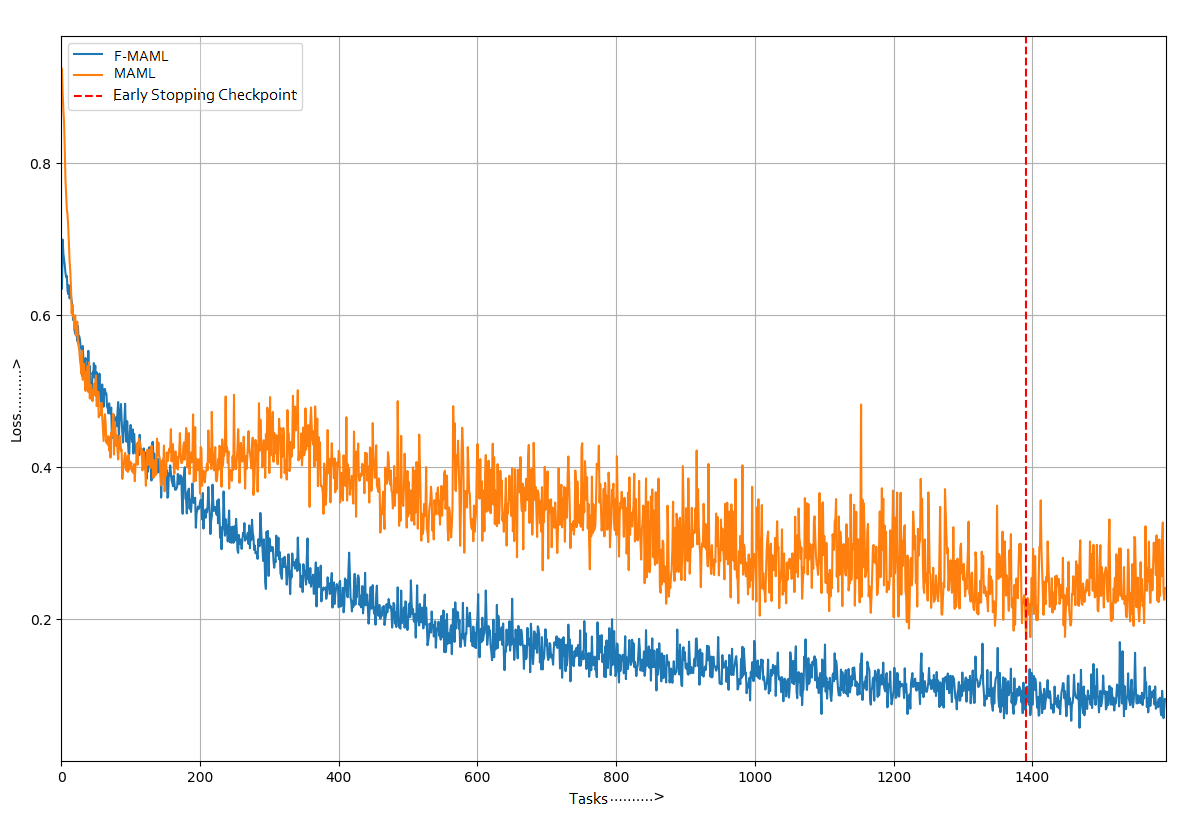}    
\caption{{Convergence Comparison MAML (MetaSER) vs F-MAML}} 
\label{fig:loss}
\end{center}
\end{figure}

\begin{table}
\begin{center}
\caption{Accuracy in 5-shot learning}\label{tb:5shot}
\begin{tabular}{c|c|c|c}
Model & English & Italian & Spanish\\\hline
Supervised & $24.33\%$ & $16.19\%$ & $20.11\%$ \\
MetaSER & $65.21\%$ & $64.13\%$ & $64.95\%$ \\ 
F-MAML & \textbf{69.71\%} & \textbf{69.13\%} & \textbf{68.85\%}\\ \hline
\end{tabular}
\end{center}
\end{table}

\begin{table}
\begin{center}
\caption{Accuracy in 10-shot learning}\label{tb:10shot}
\begin{tabular}{c|c|c|c}
Model & English & Italian & Spanish\\\hline
Supervised & $32.11\%$ & $32.54\%$ & $16.57\%$ \\
MetaSERL & $69.11\%$ & $70.13\%$ & $71.15\%$ \\ 
F-MAML & \textbf{73.71\%} & \textbf{74.22\%} & \textbf{74.55\%}\\ \hline
\end{tabular}
\end{center}
\end{table}

\begin{table}
\begin{center}
\caption{Accuracy in 20-shot learning}\label{tb:20shot}
\begin{tabular}{c|c|c|c}
Model & English & Italian & Spanish\\\hline
Supervised & $36.53\%$ & $28.64\%$ & $24.87\%$ \\
MAML & $76.21\%$ & $77.13\%$ & $77.95\%$ \\ 
F-MAML & \textbf{81.69\%} & \textbf{80.13\%} & \textbf{80.15\%}\\ \hline
\end{tabular}
\end{center}
\end{table}

Finally, our proposed approach F-MAML outperforms the MetaSER. We attribute the improvement to prior information of the fixed classes acting as anchor, which helps in efficient fine-tuning to new tasks compared to the MetaSER. The Figure-\ref{fig:loss} shows the loss of MetaSER compared to F-MAML on 5-shot learning. It can easily be seen that F-MAML converges quickly and in fewer steps than the original MAML making the proposed approach more robust that MetaSER.

\section{Conclusion}\label{sec:Conclusion}

In this paper, we simulated a scenario of SER as a few-shot learning problem. We define it as an N+F-way, K-shot problem and propose a modification to the Model-Agnostic Meta-Learning (MAML) algorithm where we kept F fixed to solve the problem. Experiments conducted on the EmoFilm dataset show that our approach performs the best compared to the baselines. In the future, we will attempt to test the feasibility of the approach on Indic languages, and mandarin derived languages since these languages differ vastly from each other. We also look for using image and text as well to make a multimodal system as well.

\bibliographystyle{IEEEtran}

\bibliography{mybib}

\end{document}